\documentclass[aps, prl, reprint, superscriptaddress, floatfix, showpacs, amsmath, amssymb]{revtex4-1}

\usepackage{lineno}
\usepackage{graphicx}
\usepackage{dcolumn}
\usepackage{bm}
\usepackage[usenames,dvipsnames,pdftex]{color} 
\usepackage{hyperref}

\setpagewiselinenumbers
\modulolinenumbers[5]

\usepackage{times}

\newcommand{\gx}{\textsc{GlueX}}

\begin{document}

\title{\boldmath Measurement of the beam asymmetry $\Sigma$ for $\pi^0$ and $\eta$ photoproduction on \\ the proton at $E_\gamma = 9$~GeV}

\affiliation{Arizona State University, Tempe, Arizona 85287, USA}
\affiliation{National and Kapodistrian University of Athens, 15771 Athens, Greece}
\affiliation{Carnegie Mellon University, Pittsburgh, Pennsylvania 15213, USA}
\affiliation{Catholic University of America, Washington, D.C. 20064, USA}
\affiliation{University of Connecticut, Storrs, Connecticut 06269, USA}
\affiliation{Florida International University, Miami, Florida 33199, USA}
\affiliation{Florida State University, Tallahassee, Florida 32306, USA}
\affiliation{The George Washington University, Washington, D.C. 20052, USA}
\affiliation{Ghent University, Proeftuinstraat 86, B-9000 Ghent, Belgium}
\affiliation{University of Glasgow, Glasgow G12 8QQ, United Kingdom}
\affiliation{GSI Helmholtzzentrum f\"ur Schwerionenforschung GmbH, D-64291 Darmstadt, Germany}
\affiliation{Indiana University, Bloomington, Indiana 47405, USA}
\affiliation{Institute for Theoretical and Experimental Physics, Moscow 117259, Russia}
\affiliation{Thomas Jefferson National Accelerator Facility, Newport News, Virginia 23606, USA}
\affiliation{University of Massachusetts, Amherst, Massachusetts 01003, USA}
\affiliation{Massachusetts Institute of Technology, Cambridge, Massachusetts 02139, USA}
\affiliation{National Research Nuclear University Moscow Engineering Physics Institute, Moscow 115409, Russia}
\affiliation{Norfolk State University, Norfolk, Virginia 23504, USA}
\affiliation{North Carolina A\&T State University, Greensboro, North Carolina 27411, USA}
\affiliation{University of North Carolina at Wilmington, Wilmington, North Carolina 28403, USA}
\affiliation{Northwestern University, Evanston, Illinois 60208, USA}
\affiliation{University of Regina, Regina, Saskatchewan, Canada S4S 0A2}
\affiliation{Universidad T\'ecnica Federico Santa Mar\'ia, Casilla 110-V Valpara\'iso, Chile}
\affiliation{Tomsk State University, 634050 Tomsk, Russia}
\affiliation{Tomsk Polytechnic University, 634050 Tomsk, Russia}
\affiliation{A. I. Alikhanian National Science Laboratory (Yerevan Physics Institute), 0036 Yerevan, Armenia}
\affiliation{College of William and Mary, Williamsburg, Virginia 23185, USA}
\affiliation{Wuhan University, Wuhan, Hubei 430072, People's Republic of China}
\author{H.~Al Ghoul}
\affiliation{Florida State University, Tallahassee, Florida 32306, USA}
\author{E.~G.~Anassontzis}
\affiliation{National and Kapodistrian University of Athens, 15771 Athens, Greece}
\author{A.~Austregesilo}
\author{F.~Barbosa}
\affiliation{Thomas Jefferson National Accelerator Facility, Newport News, Virginia 23606, USA}
\author{A.~Barnes}
\affiliation{University of Connecticut, Storrs, Connecticut 06269, USA}
\author{T.~D.~Beattie}
\affiliation{University of Regina, Regina, Saskatchewan, Canada S4S 0A2}
\author{D.~W.~Bennett}
\affiliation{Indiana University, Bloomington, Indiana 47405, USA}
\author{V.~V.~Berdnikov}
\affiliation{National Research Nuclear University Moscow Engineering Physics Institute, Moscow 115409, Russia}
\author{T.~Black}
\affiliation{University of North Carolina at Wilmington, Wilmington, North Carolina 28403, USA}
\author{W.~Boeglin}
\affiliation{Florida International University, Miami, Florida 33199, USA}
% ...
\collaboration{The \textsc{GlueX} Collaboration}
\author{V.~Mathieu}
\affiliation{Indiana University, Bloomington, Indiana 47405, USA}
\author{J.~Nys}
\affiliation{Ghent University, Proeftuinstraat 86, B-9000 Ghent, Belgium}

\pacs{13.60.Le, 13.88.+e, 14.40.Be, 12.40.Nn}


\begin{abstract} 
We report measurements of the photon beam asymmetry $\Sigma$ for the reactions $\vec{\gamma}p\to p\pi^0$ and $\vec{\gamma}p\to p\eta $ from the \gx{} experiment using a 9 GeV linearly-polarized, tagged photon beam incident on a liquid hydrogen target in Jefferson Lab's Hall D.  The asymmetries, measured as a function of the proton momentum transfer, possess greater precision than previous $\pi^0$ measurements and are the first $\eta$ measurements in this energy regime.  The results are compared with theoretical predictions based on $t$-channel, quasi-particle exchange and constrain the axial-vector component of the neutral meson production mechanism in these models.
\end{abstract}

\maketitle


In high-energy photoproduction, the dominant meson production mechanism at small momentum transfer is expected to be the exchange of massive quasi-particles known as Reggeons~\cite{Irving:1977ea}.  Interest in this theoretical description of high-energy photoproduction has increased recently, as it provides constraints on the quantum mechanical amplitudes utilized in low-energy meson photoproduction to extract the spectrum of excited baryons~\cite{Mathieu:2015gxa}, which depend strongly on the internal dynamics of the underlying constituents~\cite{CredeBaryon}.
In addition, understanding the meson photoproduction mechanism at high energies is a vital component of a broader program to search for gluonic excitations in the meson spectrum through photoproduction reactions, which is the primary goal of the \gx{} experiment at Jefferson Lab.  

The first model developed for high-energy $\vec{\gamma}p\to p\pi^0$ by Goldstein and Owens was based on the exchange of Reggeons with the allowed $t$-channel quantum numbers $J^{PC} = 1^{--}$ and $1^{+-}$, corresponding to the leading trajectories of the vector $\rho^0/\omega$ and axial-vector $b^0_1/h_1$ Reggeons, respectively, along with Regge cuts~\cite{Goldstein1973}.  Similar approaches addressing both $\pi^{0}$ and $\eta$ photoproduction have been developed and extended recently by several groups, including Laget~\cite{Laget2005,Laget:2010za}, the JPAC collaboration~\cite{Mathieu2015,Nys:2016vjz}, and Donnachie and Kalashnikova~\cite{Donnachie2016}.  Predictions for the linearly polarized beam asymmetry are sensitive to the relative contribution from vector and axial-vector exchanges, and new data can provide important constraints to better understand this production mechanism.



In this paper, we report on the linearly polarized photon beam asymmetry $\Sigma$ in high-energy $\pi^0$ and $\eta$ photoproduction from the \gx{} experiment.  The data were collected in the spring of 2016 utilizing the newly upgraded Continuous Electron Beam Accelerator Facility (CEBAF) at Jefferson Lab.  The data represent the first measurement with a 12~GeV electron beam at Jefferson Lab and the first measurement from the \gx{} experiment.  During most of this period, CEBAF provided \gx{} with a beam current of about 150~nA at a repetition rate of 250 MHz.  

The \gx{} experiment~\cite{Ghoul:2015ifw} uses a new high-energy photon beam facility, where the electrons provided by CEBAF are incident on a thin aluminum ($30~\mu$m) or diamond ($50~\mu$m) radiator, producing a tagged bremsstrahlung photon beam.  The aluminum radiator produces a conventional incoherent bremsstrahlung spectrum with the characteristic intensity proportional to $1/E_\gamma$.  The lattice structure of the diamond radiator was aligned with the beam to produce coherent bremsstrahlung, with the coherent photon intensity peaking in specific energy ranges where the photons are linearly polarized relative to the crystal axes in the diamond.  Two different diamond orientations were used for this dataset (alternating every few hours), with the electric field vector parallel or perpendicular to the floor of the experimental hall, denoted as PARA and PERP, respectively.

After passing through the thin diamond radiator, the scattered beam electrons propagate through a dipole magnet and are detected in a scintillator-hodoscope array, thus tagging the energy of the radiated beam photons.  In the photon beam energy range 3.0-11.8~GeV, there are two, independent detectors: a fine-grained Tagger Microscope instrumenting the region $8.2<E_\gamma<9.2$~GeV in increments of about 10~MeV and the Tagger Hodoscope sampling the remaining energy range with individual counter widths between 10 and 25~MeV.

\begin{figure}[!t]
\begin{center}
\includegraphics[width=0.5\textwidth]{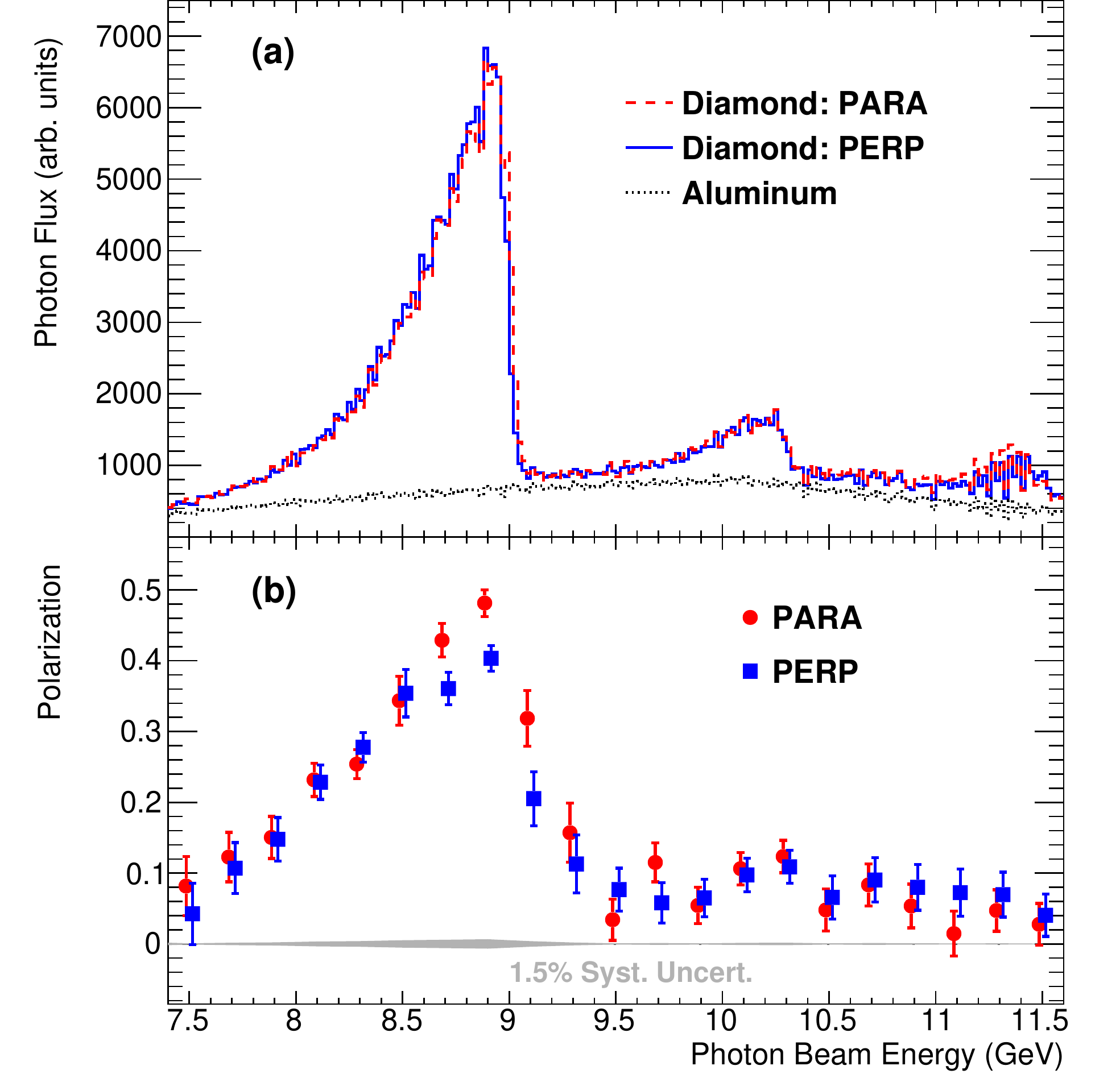}
\end{center}
 \caption {(color online) (a) Photon beam intensity versus energy as measured by the pair spectrometer (not corrected for instrumental acceptance).  (b) Photon beam polarization as a function of beam energy, as measured by the triplet polarimeter, with data points offset horizontally by $\pm0.015$~GeV for clarity.} \label{fig:beam}
\end{figure}

The beam photons are predominantly produced along the direction of the incident electron beam, with a narrower angular distribution for coherent than incoherent bremsstrahlung.  Therefore, after the photons travel through a $75$~m-long vacuum beamline, they pass through a $3.4$~mm-diameter collimator, where the off-axis photons are removed, increasing the fraction of coherently produced photons.  The energy of the photon beam is monitored using $e^+e^-$ pair conversion from a thin ($75~\mu$m) beryllium foil downstream of the collimator, where the $e^+$ and $e^-$ energies are measured in a pair spectrometer system consisting of a dipole magnet and a pair of scintillator counter arrays~\cite{Barbosa:2015bga}.  The normalized photon beam energy spectra, as measured by the pair spectrometer (not corrected for instrumental acceptance), are shown in Fig.~\ref{fig:beam}(a) for the diamond and aluminum radiators.  Here, the characteristic peak of coherent photons is clearly visible in the diamond distributions at $E_\gamma = 9$~GeV, relative to the incoherent photons from the aluminum radiator.

The polarization of the coherent photons is measured by a triplet polarimeter~\cite{Dugger:2017zoq}, where photons convert on atomic electrons in the same beryllium foil as used by the pair spectrometer, via the process $\vec{\gamma} e^- \to e^- e^+ e^-$.  The high-energy $e^+e^-$ pair is detected in the pair spectrometer, while the low energy recoil $e^-$ is detected in a 1~mm-thick silicon detector, which is segmented in azimuthal angle $\phi_{e^-}$ around the beamline.  The distribution of the recoil $e^-$ in azimuth is given by $d\sigma/d\phi_{e^-} \propto 1 + P_\gamma\lambda \cos2(\phi_{e^-} - \phi_\gamma^{\rm{lin}})$, where $P_\gamma$ is the photon beam linear polarization, $\phi_\gamma^{\rm{lin}}$ is the azimuthal angle of the beam photon's linear polarization plane, and $\lambda$ is the analyzing power, which is calculable in QED.

The linear polarization is extracted from the measured $\phi_{e^-}$ distribution for both PARA ($\phi_\gamma^{\rm{lin}}=0^\circ$) and PERP ($\phi_\gamma^{\rm{lin}}=90^\circ$) configurations and peaks with the coherent photon intensity at $E_\gamma = 9$~GeV as shown in Fig.~\ref{fig:beam}(b).   The polarization was weighted by the beam energy distribution for reconstructed $\vec{\gamma} p \to p\pi^0$ events to determine the average value in the energy range $8.4 < E_{\gamma} < 9.0$ GeV: $P^{\rm{PARA}}_\gamma=0.440 \pm 0.009 \rm{(stat.)} \pm 0.007 \rm{(syst.)}$ and $P^{\rm{PERP}}_\gamma=0.382 \pm 0.008 \rm{(stat.)} \pm 0.006 \rm{(syst.)}$.  The statistical uncertainties of 2.1\% are independent for both polarizations and driven by the yield of triplet production events in the data sample.  The correlated systematic uncertainty inherent in the design and operation of the triplet polarimeter is 1.5\%, as documented in Ref.~\cite{Dugger:2017zoq}.  

The statistical precision of our dataset prohibits us from probing additional systematic uncertainties on the beam polarization below the 2\% level, and we have no evidence of additional systematic errors at or above this level.  Therefore, considering the independent statistical errors on the polarization measurements and the systematic error of 1.5\%, we assume a total error of 2.1\% on the sum of the polarizations, which normalize the extracted beam asymmetry in Eqn.~\ref{eqn:asym}.  This uncertainty is fully correlated between the $\vec{\gamma}p\to p\pi^0$ and $\vec{\gamma}p\to p\eta$ reactions.

The difference between the measured polarizations for the two configurations is consistent with independent fits to the observed azimuthal asymmetry for $\vec{\gamma} p \rightarrow p\pi^0$ events for PARA and PERP separately, and may be due to different electron beam positions on the diamond or different collimation conditions for the PARA and PERP configurations.
The integrated luminosity of the dataset used in this analysis is approximately 1~pb$^{-1}$ in the coherent-peak energy range.


The \gx{} experiment is a large-acceptance, azimuthally symmetric detector for both charged particles and photons.  It is located in the recently constructed experimental Hall D at Jefferson Lab.  The central region of \gx{} is contained within a solenoid magnet, which provides a 1.8~T magnetic field along the direction of the beam.  The collimated photon beam is incident on a $30$~cm-long unpolarized, liquid hydrogen target located 1.3~m upstream of the solenoid's center.  Surrounding the target is the Start Counter, a segmented cylindrical scintillator detector with a cone section that tapers towards the beamline on the downstream end, which provides a measure of the primary interaction time with a resolution of better than 300~ps.

The Central Drift Chamber (CDC)~\cite{VanHaarlem:2010yq} is located just outside the Start Counter and contains $28$ layers of straw tubes, including axial and stereo layers, which are $150$~cm in length and located radially between $10$~cm and $59$~cm.  Downstream of the CDC there are four packages of the Forward Drift Chamber (FDC)~\cite{Berdnikov2015}, which stretch 2~m along the beamline.  Each package is based on six layers of planar drift chambers with both anode and cathode readouts, providing three-dimensional space points.  In combination, the CDC and FDC provide charged-particle tracking with uniform azimuthal coverage over polar angles $1^{\circ}-120^{\circ}$.

Surrounding the tracking devices inside the solenoid is the Barrel Calorimeter (BCAL)~\cite{Leverington:2008zz,Smith:2016yvt}, which covers polar angles between $12^{\circ}$ and $120^{\circ}$. The BCAL is a lead-scintillating fiber calorimeter with readout on both the upstream and downstream ends.  The Forward Calorimeter (FCAL)~\cite{Moriya:2013aja} is located ${\sim}6$~m downstream of the target and consists of 2,800 lead-glass blocks oriented such that the FCAL acceptance is azimuthally symmetric for polar angles $1^{\circ}-11^{\circ}$.  The detector readout was triggered by a significant energy deposit in the BCAL or FCAL.


\begin{figure}[!t]
\begin{center}
\includegraphics[width=0.5\textwidth]{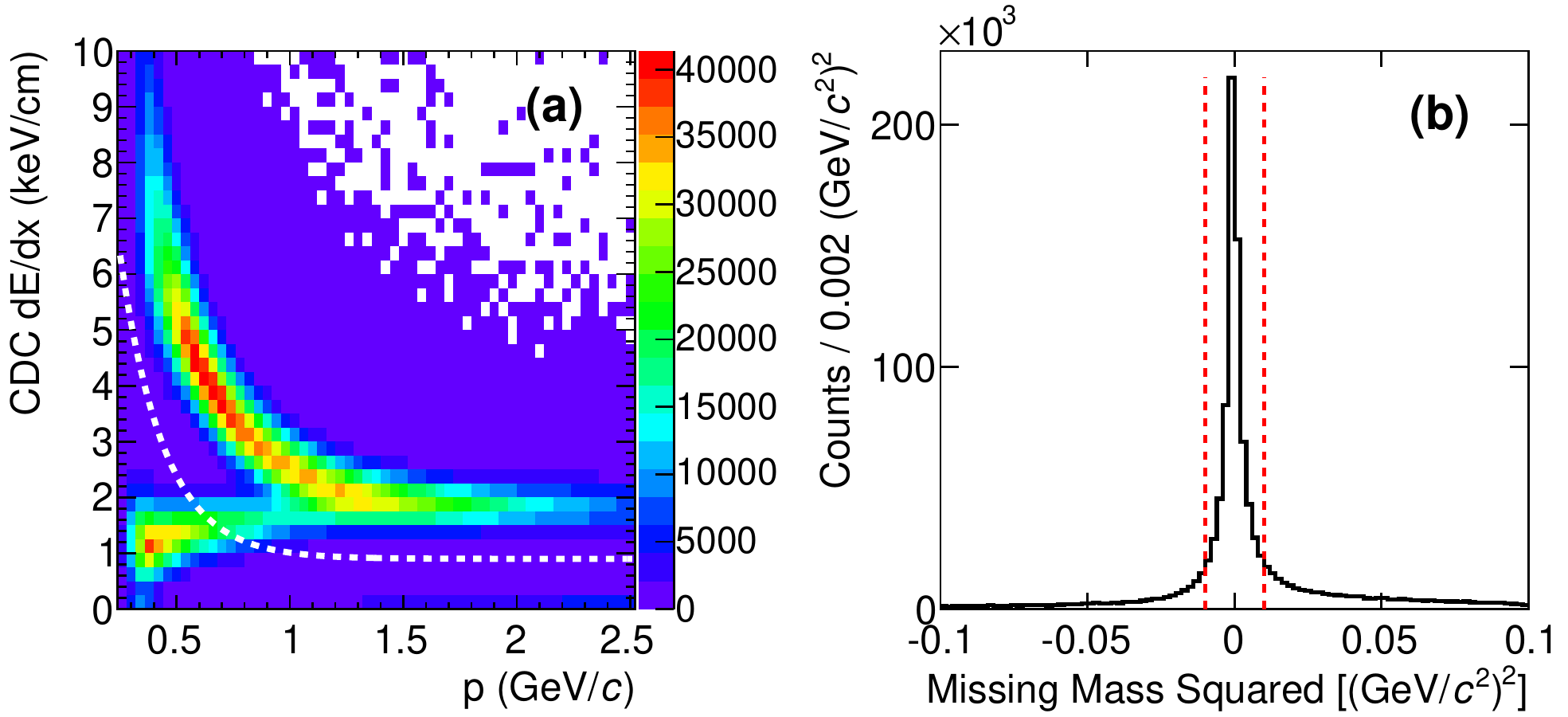}
\end{center}
 \caption {(color online) (a) Energy loss $dE/dx$ versus postively charged track momentum and (b) the spectrum of missing mass squared for the reaction $\vec{\gamma} p \to p\gamma\gamma$.} \label{fig:cuts}
\end{figure}

The $\pi^0 p$ and $\eta p$ final states were detected through the $\pi^0 \to \gamma\gamma$ and $\eta \to \gamma\gamma$ decay modes.  The selection of exclusive $\vec{\gamma} p \to p\gamma\gamma$ events began by identifying all events with at least the following: one tagged beam photon, one positively charged track with $p>0.25$~GeV$/c$ originating from the target region, and two neutral showers in the calorimeters.  The time of the primary interaction was determined by a Start Counter hit matched to the proton track, which identifies the Radio Frequency (RF) bunch of the electron beam.  The time difference $\Delta t = t_{\rm{beam}} - t_{\rm{RF}}$ between the tagged beam photon and the machine RF signal was then used to select tagged beam photons that were associated with the primary interaction by requiring $|\Delta t| < 2$~ns.  To account for the tagged photons that were accidentally associated with the RF bunch of the primary interaction, we selected a separate sample of events, referred to as ``accidentals," where $6<|\Delta t|<18$~ns.  This ``accidentals" sample (scaled by a factor of 1/6) was used to statistically subtract the contribution of the accidentally tagged photons from the primary RF bunch.

The vast majority of the proton candidate tracks traverse the CDC, which, in addition to providing spatial points for the track reconstruction, also provides a measure of the energy loss $dE/dx$ for charged particles.  Figure~\ref{fig:cuts}(a) shows the energy loss versus momentum for the proton candidate tracks, where a clear separation between protons and pions is observed for momenta less than 1~GeV$/c$.  Protons were selected by requiring a measured $dE/dx$ greater than the dashed white curve in Fig.~\ref{fig:cuts}(a).

The exclusive nature of the $\vec{\gamma} p \to p\pi^0$ and $\vec{\gamma} p \to p\eta$ reactions provides kinematic constraints on the measured particles, as both the initial beam energy and the momenta of all the final-state particles are measured in \gx{}.  Thus, the requirements that energy and momentum are conserved in the interaction allows for a strong rejection of background processes in the selection of events.  Considering only the final-state particles, the transverse momentum balance was studied by reconstructing the azimuthal angle difference $\Delta\phi = \phi_{p} - \phi_{\gamma\gamma}$ and requiring $|\Delta\phi - 180^\circ| < 5^\circ$.

\begin{figure}[!t]
\begin{center}
\includegraphics[width=0.5\textwidth]{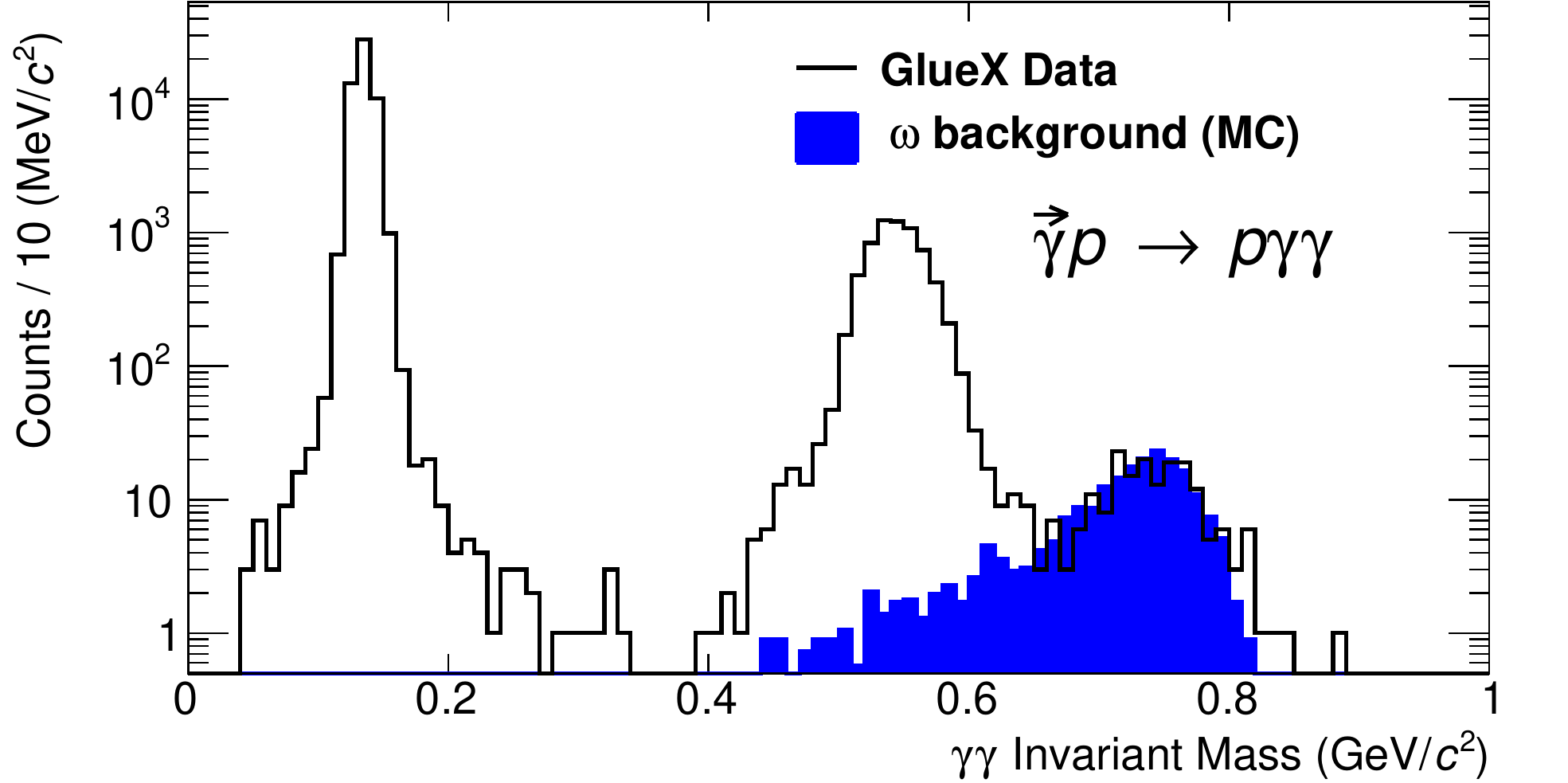}
\end{center}
 \caption {(color online) $\gamma\gamma$ invariant mass distribution with clear peaks at the $\pi^0$ and $\eta$ masses, superimposed with background estimated from $\vec{\gamma} p \to p\omega$, $\omega \to \pi^0\gamma$ simulation.} \label{fig:m2g}
\end{figure}

To reduce contributions from processes with additional massive particle(s) not detected in the final state, we considered the missing mass for the signal reaction $\vec{\gamma} p \to p\gamma\gamma$.  The missing mass is defined as the magnitude of the 4-momentum difference between the initial and final state particles.  The square of the missing mass is shown with ``accidentals" subtracted in Fig.~\ref{fig:cuts}(b) and the absolute value was required to be less than 0.01~(GeV$/c^2)^2$, as shown by the red dashed lines.  Also, the missing energy ($\Delta E$) was required to be $-0.5 < \Delta E < 0.7$~GeV to eliminate reactions with a missing photon.  


The process $\vec{\gamma} p \to p\omega$, $\omega \to \pi^0\gamma$ contributes background to the $p\eta$ final state, where a low-energy photon from the $\pi^0$ goes undetected.  To reduce this background, the exclusive kinematics were again used to provide a constraint on the missing mass in the reaction $\vec{\gamma} p \to pX$, where the final-state photons are treated as missing.  The missing mass $M_X$ requirement was: $M_X <$ 0.5 (0.7)~GeV$/c^2$ for the $\pi^0$ ($\eta$) reaction. 
 
As a final constraint on the exclusivity of the reaction, the sum of the energies from all of the BCAL and FCAL calorimeter hits in the event was computed, excluding those hits corresponding to the reconstructed photons from the $\pi^0$ or $\eta$ decay and those associated with the reconstructed proton track.  Any excess energy in this sum would be due to additional particles in the final state.  These events were rejected by requiring the excess energy to be less than 17~MeV, as set by the low energy sensitivity for the BCAL.  Finally, the beam photon energy range $8.4<E_\gamma<9.0$~GeV was selected to enhance the contribution from linearly polarized photons.  

The candidates surviving the described event selection are shown in Fig.~\ref{fig:m2g} as a function of the invariant mass of the two photons, with the $y$-axis given in logarithmic scale.  Clear peaks are observed at the $\pi^0$ and $\eta$ masses, with Gaussian widths $\sigma=7$ and 21~MeV$/c^2$, respectively.  The $\vec{\gamma} p \to p\pi^0$ and $\vec{\gamma} p \to p\eta$ candidate events were selected by requiring the measured $M_{\gamma\gamma}$ to be within $\pm3\sigma$ of the known masses.  Phase-space Monte Carlo (MC) events for the process $\vec{\gamma} p \to p\omega$ were generated, passed through a GEANT 3~\cite{Brun:1978fy} model of the \gx{} detector, and subjected to the same event-selection criteria as the data.  The surviving $\omega$ background sample is shown in Fig.~\ref{fig:m2g}, normalized to the data in the $\omega$ mass range.  After all the event criteria were applied, the $\omega$ background contribution in the $\eta$ mass range was ${\sim}0.38$\%, and the contribution to the $\pi^0$ yield was negligible.

\begin{figure}[!t]
\begin{center}
\includegraphics[width=0.5\textwidth]{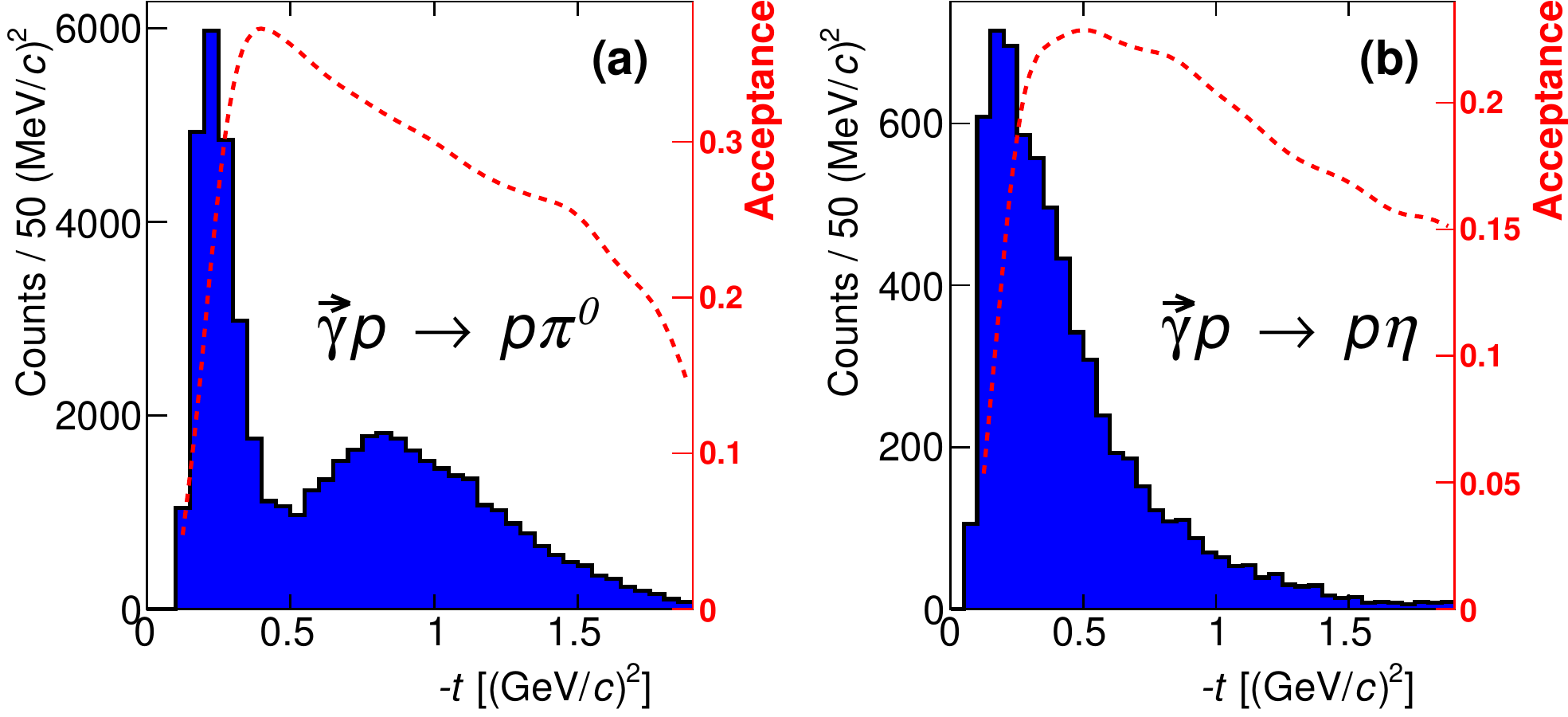}
\end{center}
 \caption {(color online) Candidate event yield as a function of the proton momentum transfer $-t$ for (a) $\vec{\gamma} p \to p\pi^0$ and (b) $\vec{\gamma} p \to p\eta$, without corrections for instrumental acceptance.  The acceptance functions (red dashed), determined from MC simulation, are shown for comparison.} \label{fig:t}
\end{figure}

\begin{figure*}[]
\begin{center}
\includegraphics[width=1.0\textwidth]{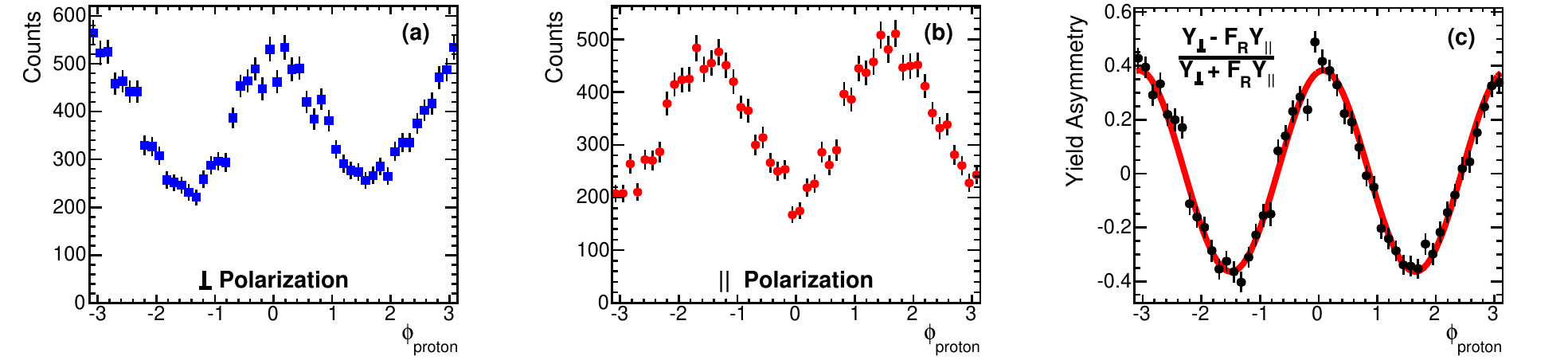}
\end{center}
 \caption {(color online) $\vec{\gamma}p\to p\pi^0$ yield (statistical errors only) versus $\phi_p$ integrated over $-t$ for (a) PERP and (b) PARA.  (c) The yield asymmetry, fit with Eq.~(\ref{eqn:asym}) to extract $\Sigma$.} \label{fig:asym}
\end{figure*}


Figure~\ref{fig:t} shows the $\pi^0$ and $\eta$ yields (without corrections for instrumental acceptance) as a function of the proton momentum transfer $t=(p_{\rm{target}} - p_{\rm{p}})^2$.  The acceptance functions in Fig.~\ref{fig:t} were determined from MC simulation utilizing Regge models~\cite{Mathieu2015,Laget2005}, and do not significantly alter the distributions apart from the threshold at low $-t$.  The $\vec{\gamma} p \to p\pi^0$ distribution shows the expected dip near $-t = 0.5$~(GeV$/c)^2$ observed in previous measurements~\cite{Anderson1971}, which is characteristic of a zero in the dominant $\omega$ Reggeon exchange.  The $\vec{\gamma} p \to p\eta$ distribution does not show a dip in the observed $-t$ range, also consistent with previous measurements~\cite{Anderson:1969kn}.


The azimuthal dependence of the cross section for the photoproduction of pseudoscalar mesons with a linearly polarized photon beam and an unpolarized target is given by:
\begin{equation} 
\sigma = \sigma_0 \left( 1 - P_\gamma\Sigma \cos2(\phi_p - \phi_\gamma^{\rm{lin}}) \right),
\end{equation}
\noindent where $\sigma_0$ is the unpolarized cross section, $\Sigma$ is the linearly polarized beam asymmetry, and $\phi_p$ is the azimuthal angle of the production plane defined by the final-state proton~\cite{Barker:1975bp}.  Therefore, the yields for the PERP and PARA orientations are given by:
\begin{eqnarray} 
Y_\perp &\propto N_\perp(1 + P_\perp\Sigma \cos2\phi_p) \\
Y_\parallel &\propto N_\parallel(1 - P_\parallel\Sigma \cos2\phi_p),
\end{eqnarray}
\noindent and are shown in Fig.~\ref{fig:asym} (a) and (b), respectively, integrated over all $t$ after subtracting the background contribution from accidentally tagged photons.  The azimuthal symmetry of the \gx{} detector provides a clear visualization of the $1 \pm P_\gamma\Sigma \cos2\phi_p$ dependence of the yield without any correction for instrumental acceptance.

The orthogonality of the PARA and PERP polarization configurations provides an exact cancelation of any $\phi$-dependent instrumental acceptance through a measurement of the yield asymmetry
\begin{equation}
\label{eqn:asym} 
\frac{Y_\perp - F_R Y_\parallel}{Y_\perp + F_R Y_\parallel} = \frac{(P_\perp + P_\parallel)\Sigma \cos2\phi_p}{2 + (P_\perp - P_\parallel)\Sigma \cos2\phi_p},
\end{equation}
where $F_R = N_\perp/N_\parallel$ is the ratio of the integrated photon flux between PERP $(N_\perp)$ and PARA $(N_\parallel)$.  The flux ratio was determined to be $F_R = 1.04 \pm 0.05$ by integrating the yield of coincidences between the pair spectrometer and tagger microscope for each beam orientation.  Figure~\ref{fig:asym}(c) shows the yield asymmetry as a function of $\phi_p$, which is fit using the functional form in Eq.~(\ref{eqn:asym}), where the only free parameter is the beam asymmetry $\Sigma$.

\begin{figure}[!ht]
\begin{center}
\includegraphics[width=0.5\textwidth]{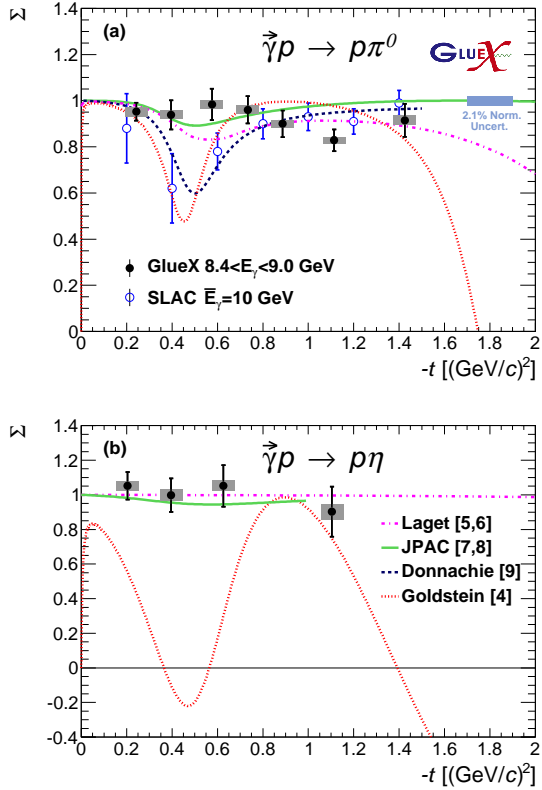}
\end{center}
 \caption {(color online) Beam asymmetry $\Sigma$ for (a) $\vec{\gamma}p\to p\pi^0$ and (b) $\vec{\gamma}p\to p\eta$ (black filled circles).  Uncorrelated systematic errors are indicated by gray bars and combined statistical and systematic uncertainties are given by the black error bars. The previous SLAC results~\cite{Anderson1971} at $\overline{E}_\gamma=$10~GeV (blue open circles) are also shown along with various Regge theory calculations.} \label{fig:sigma}
\end{figure}

Following the procedure described above to extract $\Sigma$, the yield asymmetry is determined in bins of $-t$ for the $\pi^0$ and $\eta$ reactions, for which the results are shown in Fig.~\ref{fig:sigma}~\cite{supplement}.  Systematic uncertainties due to the event selection were determined by measuring the asymmetries in each $-t$ bin with varied selection criteria and resulted in uncertainties of 1-2\% for $\pi^0$ and 2-4\% for $\eta$.  The flux ratio uncertainty contributes $1\%$ to the measured asymmetries, and a $1\%$ uncertainty was estimated for the $\omega$ background contribution to the $\eta$ sample.  The asymmetries have a common $2.1\%$ normalization uncertainty due to the beam polarization.

Several Regge theory calculations for the beam asymmetries at $E_\gamma=9$~GeV are shown in Fig.~\ref{fig:sigma} for comparison~\cite{Donnachie2016,Goldstein1973,Mathieu2015,Nys:2016vjz,Laget2005,Laget:2010za}.  
Some of these calculations incorporate a significant dip in the asymmetries near $-t = 0.5$~(GeV$/c)^2$, due to a contribution from the axial-vector Reggeon exchange that is consistent with previous $\pi^0$ measurements at $\overline{E}_\gamma=10$~GeV from SLAC~\cite{Anderson1971}.  This dip is not observed in the \gx{} data, which indicates a dominance of the vector Reggeon exchange at this energy.

In summary, we report on the linearly polarized photon beam asymmetry $\Sigma$ for $\vec{\gamma}p\to p\pi^0$ and $\vec{\gamma}p\to p\eta$ by the \gx{} experiment at $E_\gamma=9$~GeV and $0.15<-t<1.6$~(GeV$/c)^2$.  These are the first measurements utilizing the 12~GeV electron beam and the new, high-energy photon beam facility in Hall D at Jefferson Lab, opening a new era in the study of polarized photoproduction.  The results for the $\pi^0$ asymmetry represent a significant increase in precision relative to previous measurements, and the $\eta$ measurements are the first above $E_\gamma = 3$~GeV.  The asymmetries are compared to existing Regge calculations and are expected to contribute to our understanding of production mechanisms in high-energy photoproduction necessary to search for exotic meson states with future high-statistics data samples. 

\begin{acknowledgments}
We would like to acknowledge the outstanding efforts of the staff of the Accelerator and the Physics Divisions at Jefferson Lab that made the experiment possible.  We acknowledge Alex Dzierba for his essential contributions to the conception and development of the \gx{} experiment and his leadership of the \gx{} Collaboration. We appreciate communication with Alexander Donnachie, Gary Goldstein, Yulia Kalashnikova, Jean-Marc Laget and the JPAC collaboration.  
This work was supported in part by the U.S. Department of Energy, the U.S. National Science Foundation, the Natural Sciences and Engineering Research Council of Canada, the Russian Foundation for Basic Research, the UK Science and Technology Facilities Council, the Chilean Comisi\'{o}n Nacional de Investigaci\'{o}n Cient\'{i}fica y Tecnol\'{o}gica, the National Natural Science Foundation of China, and the China Scholarship Council.  This material is based upon work supported by the U.S. Department of Energy, Office of Science, Office of Nuclear Physics under contract DE-AC05-06OR23177.
\end{acknowledgments}

\bibliography{pseudoBeamAsym2016}

\end{document}